# Mode- and size-dependent Landau-Lifshitz damping in magnetic nanostructures: Evidence for non-local damping

Hans T. Nembach, Justin M. Shaw, Carl T. Boone, T. J. Silva
Electromagnetics Division, National Institute of Standards and Technology, Boulder, CO 80305, USA

Abstract: We demonstrate a strong dependence of the effective damping on the nanomagnet size and the particular spin-wave mode that can be explained by the theory of intralayer transverse-spin-pumping. The effective Landau-Lifshitz damping is measured optically in individual, isolated nanomagnets as small as 100 nm. The measurements are accomplished by use of a novel heterodyne magneto-optical microwave microscope with unprecedented sensitivity. Experimental data reveal multiple standing spin-wave modes that we identify by use of micromagnetic modeling as having either localized or delocalized character, described generically as end- and center-modes. The damping parameter of the two modes depends on both the size of the nanomagnet as well as the particular spin-wave mode that is excited, with values that are enhanced by as much as 40% relative to that measured for an extended film. Contrary to expectations based on the *ad hoc* consideration of lithography-induced edge damage, the damping for the end-mode decreases as the size of the nanomagnet decreases. The data agree with the theory for damping caused by the flow of intralayer transverse spin-currents driven by the magnetization curvature. These results have serious implications for the performance of nanoscale spintronic devices such as spin-torque-transfer magnetic random access memory.

The Landau-Lifshitz and Gilbert equations [1] [2] [3], both with purely local formulations of the damping term, are regarded as the definitive phenomenological descriptions of dissipative ferromagnetic dynamics. Most micromagnetic simulations for magnetization dynamics rely on the local damping formulation in a diverse variety of systems, e.g. disk drives [4], telecommunications [5], and biomolecule sorting [6]. However, an outstanding question is damped gyromagnetic precession subject to finite size effects at the nanometer scale: Should one expect damping to be identical for a 10 nm and a 10 cm body, all else being equal? The answer to this question is of great technological significance for a broad range of applications. For example, the damping parameter $\alpha$ is a critical figure-of-merit for the efficient operation of many spintronic devices, e.g., spin-torque-transfer magnetic random access memory (STT-MRAM) devices that are potentially scalable down to the 22 nm lithography node and beyond [7]. In the case of STT-MRAM, the switching energy scales quadratically with switching current, which is in turn proportional to $\alpha$; thus, small $\alpha$ is essential for low power operation.

The leading theory for damping in ferromagnetic conductors is magnon-electron scattering [8] [9], whereby intrinsic damping is purely local at room temperature [10]. To date, spin-pumping, which drives spin-current from a ferromagnet into adjacent non-magnetic conducting layers, is the only experimentally confirmed mechanism of extrinsic nonlocal damping [11]. Recent theoretical work describes *intrinsic* nonlocal damping due to the dissipative flow of non-equilibrium *intralayer* spin-currents within the ferromagnet itself [12] [13] [14] [15], which can give rise to enhanced damping in isolated magnetic nanostructures. Evidence in support of such theories remains inconclusive. Experimentally, spin-torque ferromagnetic resonance (ST-FMR) has been widely used to measure damping in individual nanoscale devices. While the damping is often found to be larger than values reported for extended thin films (measured damping values for Permalloy in nanopillars by use of ST-FMR range from 0.010±0.002 at room temperature [16] to 0.016 at 4.2 K [17]. The intrinsic $\alpha$ for thin film Permalloy is only 0.004 +/- 0.001 [18],) this discrepancy has often been attributed to increased damping close to the edges of the nanomagnets, the result of damage, re-deposition and/or oxidation at the sidewalls [17]. Unfortunately, the interpretation of ST-FMR data is made difficult by the complexity of the multilayer structures, Oersted field effects, and the difficulty in isolating the contributions to damping from interlayer interactions. We now demonstrate that *intrinsic* non-local effects, moderated by spin-wave mode confinement, are important contributors to damping in magnetic nanostructures. Indeed, we show that both interlayer *and intralayer* spin-pumping are of comparable magnitude for the nanoscale systems considered here.

Our approach is to measure the dynamics in individual nanomagnets with a single ferromagnetic layer. This allows determination of the intrinsic properties of the quantized spin-wave modes without influence of other adjacent ferromagnetic layers. Extraction of $\alpha$ from ensemble measurements of nanomagnet arrays is not trivial, both because (a) the resonance frequencies might differ from nanomagnet to nanomagnet [19] [20], and (b) shape distortions can give rise to mode splitting [21], both sources of extrinsic linewidth broadening. Therefore, measurement of the linewidth of individual nanomagnets is essential. In addition, a more systematic comparison of data with theory is made possible by examination of the dependence of damping on various spin-wave modes in nanomagnets of differing size [12] [15] [13] [14].

Measurement of $\alpha$ in individual nanomagnets has been achieved with the time-resolved magneto-optical Kerr effect (MOKE) [22] [23] [24], but such measurements are challenging when the diffraction-limited spot-size for focused visible light is much larger than the nanomagnet, adversely affecting the signal-to-noise-ratio (SNR). The SNR of weak optical signals can be enhanced by use of optical heterodyne detection, where the optical signal is mixed with a bright local oscillator (LO) beam [25]. We developed a novel heterodyne magneto-optical microwave microscope (H-MOMM) to measure ferromagnetic resonance (FMR) in individual, well-separated nanomagnets by use of heterodyne detection of magneto-optical signals at microwave frequencies. The signal from a spin-wave mode, e.g. the end-modes in the 200 nm nanomagnets, which are localized in an $\approx 2100\,\text{nm}^2$ area, measured with the H-MOMM is more than 10 times larger than measured with a conventional magneto-optical Kerr microscope. (See SI.)

Samples were prepared from thin films of 3 nm Ta/10 nm $Ni_{80}Fe_{20}$/5 nm $Si_3N_4$ on 100-μm-thick sapphire substrates. Elliptical-shaped nanomagnets with nominal dimensions of 480×400 $nm^2$, 240×200 $nm^2$ and 120×100 $nm^2$ were patterned by e-beam/ion-mill lithography. 20×20 $\mu m^2$ squares were also patterned from the same films to facilitate determination of the blanket-film FMR properties (See Ref. [21] for details).

FMR spectra for two of the 400 nm nanomagnets, and three each of the 200 nm and 100 nm nanomagnets, were measured over a wide frequency range. The spectra were obtained by fixed frequency excitation and by sweeping the external magnetic field $H_{ext}$ that was applied along the nanomagnet long axis. The microwave field from the waveguide was oriented along the short axis. The inset in Fig. 1 shows an example of a 13.2 GHz spectrum with a 100 nm magnet. As was previously demonstrated in Ref. [20], we also compared our data to micromagnetic simulations to confirm the identity of the various resonances as being associated with end- and center-mode excitations. The identification was both qualitatively and quantitatively conclusive. Further comparison of the data with micromagnetic simulations (described below) indicate that the spin-wave mode with the lowest resonance field (i.e., the "center mode") is distributed throughout the volume of the nanomagnet, and the two other modes (i.e. the "end-modes") are localized at the ends of the nanomagnet along the applied field direction [20]. A perfect elliptical nanomagnet would have degenerate end modes, but shape distortions can lift this degeneracy, as was recently demonstrated in BLS measurements [21]. Coupling between the end modes can also break the degeneracy, but this was determined to be negligible for the systems studied here, as discussed below.

The measured amplitudes of the end modes in the 100 nm nanomagnet are significantly larger than that of the center mode. Micromagnetic simulations (see insets in Fig. 2) indicate that the center mode actually has significant amplitude at two ends of the nanomagnet, but the precession is 180° out-of-phase with respect to the central part of the mode. The heterodyne signals from the center and ends have opposite signs, which leads to partial destructive interference. Additional simulations confirm that the integrated H-MOMM signals from central and end portions of the center mode for the 100 nm nanomagnet should be comparable in magnitude, which explains the weak heterodyne signal from the center mode.

The measured magnitude spectra were fitted with the magnitude of the complex susceptibility $\chi_{xy}$ [26] (red line in inset of Fig. 1), see the SI. The resonance field

$H_{res}^{(i)}(f)$ for each mode was then fitted with the Kittel equation to extract global values for $H_1^{(i)}$ and $H_2^{(i)}$:

$$f = \left(\frac{|\gamma|\mu_0}{2\pi}\right)\sqrt{\left[H_{res}^{(i)}(f) + H_1^{(i)}\right]\left[H_{res}^{(i)}(f) + H_2^{(i)}\right]}. \quad (1)$$

The fits of the resonance field to the frequency for the center and the two end modes for a 100 nm nanomagnet are shown in Fig. 1. The center mode has a lower resonance field and less curvature than the two end-modes, while the frequency-dependence of the two end-modes is virtually identical except for a fixed field splitting of $\approx 25$ mT for the 100 nm nanomagnets.

In the case where the two end-modes are not degenerate but are coupled due to magnetostatic interactions, one might expect that modes with optical and acoustic character are excited. We used micromagnetic simulations to determine the coupling between the end-modes for the 100 nm nanomagnet. Simulations yielded a mode splitting of 5 mT at 10 GHz. Appealing to a classical model of coupled, lossy harmonic oscillators [27], the effective coupling strength between two end modes is calculated to be 28 mT. Such a coupling strength is close to the experimentally observed spitting of 25 mT for the two end-modes in Fig. 1. This implies that the measured modes are not purely localized at either of the two ends, but instead have a degree of mixed even- or odd-like character, with the excitation of one end-mode necessarily driving the other end mode with a fractional amplitude of $\approx 0.08$. We interpret the high field peak to be the odd-like mode, and the low field peak to be the even-like mode.

The fits of the spectra also yield the frequency-dependence of the linewidth for each spin-wave mode. The linewidth of a localized spin-wave mode for a single nanomagnet does not have any contributions from inhomogeneous linewidth broadening $\Delta H_0$ because the resonance frequency is necessarily homogenous for a single eigenmode. Moreover, extrapolation of the H-MOMM-measured linewidth data for the 20×20 μm² square resulted in $\mu_0 \Delta H \approx 0$ mT at $f = 0$. Thus, we can safely fit the linewidths with

$$\Delta H = (4\pi\alpha f)/(|\gamma|\mu_0). \quad (2)$$

Using eq. (2), we extracted $\alpha = 0.0074 \pm 0.0001$ for the 20×20 μm² square. This value is larger than the previously reported value of 0.004 in Ref. [18]. We attribute most of the discrepancy to spin pumping at the $Ni_{80}Fe_{20}/Ta$ interface [28] [29] [30]. To determine the spin-mixing conductance, we measured nearly identical, unpatterned $Ni_{80}Fe_{20}/Ta$ films with thicknesses varying from 5 nm to 20 nm by broadband perpendicular FMR. This geometry eliminates two-magnon scattering for the unpatterned film [31]. The asymptotic intrinsic damping is $\alpha = 0.0050 \pm 0.0001$, in good agreement with the theoretical value α=0.0046 [32], and the spin-mixing conductance is $(1.48 \pm 0.05) \times 10^{19}$ m$^{-2}$. Based on these values, the predicted damping for a 10 nm film is $0.0079 \pm 0.0002$, in reasonable agreement with our optically measured value for the 20×20 μm² square. Given this agreement, we exclude two-magnon scattering as a significant source of linewidth for the optical measurements.

The measured linewidth for the nanomagnets does not exhibit a linear dependence on frequency at the lowest frequencies. This is understood because the magnetization distribution is not uniform at low applied fields. The dipolar fields near the ends of the nanomagnet are highly nonuniform, thereby inducing an inhomogeneous magnetization configuration if the applied fields are less than or equal to the dipolar fields. Such a change of the magnetization distribution also causes the resonance field for a particular excitation frequency to decrease with decreasing field. This "field-dragging" effect leads to a distortion of the resonance curve, which results in an anomalous increase in the linewidth at low frequencies. Micromagnetic simulations confirmed this behavior. To minimize the influence of the field-dragging effect on the experimentally determined $\alpha$, we use a low frequency cut-off to restrict the range of linewidth data fitted to eq. (2). (The cut-off frequency is determined by minimizing the rms error between the data and the linear fit.) Fig. 2 shows the dependence of Δ$H$ on $f$ for the center-mode and one of the end-modes for a 200 nm and a 100 nm nanomagnet. The solid black lines are fits to eq. (2).

The average values of $\alpha$ for the center- and end-modes for a sample of three 100 nm, three 200 nm, and two 400 nm nanomagnets are plotted in Fig. 3 as a function of sample size. For reference, the value of $\alpha$ for the 20×20 μm² square is shown as a thick purple line, where the estimated error in the fitted value is the width of the line. (See the SI for the α values of all measured nanomagnets.) Of particular note, $\alpha$ for the end-mode *decreases* by almost 30% as the size of the nanomagnet is reduced from 400 nm to 100 nm, in stark contrast to what had been observed previously for the ensemble behavior of large nanomagnet arrays, where the end-mode damping *increased* by 20% as the nanomagnet size in the array was reduced from 200 to 100 nm [20]. This highlights the advantage of the H-MOMM technique, whereby we can now extract the damping properties of individual structures without any obscuration due to structure-to-structure variations, which can otherwise complicate the process of extricating intrinsic damping from inhomogenous broadening effects [21].

There are several different models that might explain the dependence of damping on nanomagnet size. By comparing the measured size dependence of the extracted damping to that predicted for each of the models, we show that only an increase due to non-local damping resulting from intralayer dissipative transverse spin-currents is consistent with the experimental data. We explicitly show that damage and/or oxidation at the sidewalls of the nanomagnets cannot explain the experimental data.

Previous work [13] has predicted that longitudinal [14] and transverse [15] *intralayer* spin-currents can increase the damping when the dynamics are spatially inhomogenous. The net damping torque density is given by

$$\vec{T}_{damp} = -(\alpha M_s/|\gamma|)(\vec{m} \times \partial_t \vec{m}) + (\sigma_T \vec{m} \times \nabla^2 \partial_t \vec{m}), \qquad (3)$$

where $\sigma_T = (h/2)^2 n_e \tau_{sc}/m^*$ is the transverse spin conductivity, $n_e$ is the conduction electron density, $m^*$ the effective mass, and $\tau_{sc}$ is the transverse spin scattering time, which can have contributions from momentum scattering, e-e interactions, as well as spin-orbit induced spin-flip/decoherence processes. The Laplacian operator in eq. (3) implies that the damping for a given Fourier component of a localized spin-wave mode is proportional to the square of the wavenumber. Assuming that the net damping of a given eigenmode is determined by the integral of the Laplacian for the mode, normalized by the mode area, we can use simulated mode profiles from micromagnetics to estimate the enhanced damping due to intralayer spin-currents. In Fig. 3(a) we show the measured $\alpha$ and in (b) the best fit of the data, with the result $\tau_{sc}$ = 49 fs as the sole fitting parameter (See the SI for details). We use $n_e = k_F^3/3\pi^2 = 3.9 \cdot 10^{28}$ m$^{-3}$ from the measured Fermi wavenumber $k_F = 1.05 \cdot 10^{10}$ m$^{-1}$ for the majority band in Permalloy [33] and the free electron mass for $m^*$.

The theory of non-local damping due to intralayer spin-currents provides an intuitively appealing explanation for the decrease in damping observed for the end-modes when the nanomagnet size is reduced from 200 nm to 100 nm. As the size of the nanomagnet shrinks, the two localized modes on opposite ends of the nanomagnet merge together. In doing so, as seen in the insets of Fig. 2, the combined mode becomes more uniform; thereby decreasing the components of damping that are proportional to $k^2$. However, micromagnetic simulations also show that the opposite is true of the center mode; shrinking the nanomagnet "squeezes" the mode structure into a smaller area, causing the mode profile to become less uniform, with the final result that the damping increases with decreasing spatial dimension.

Based on reported values for the spin diffusion length of l$_{sf}$ = 3 nm - 8 nm [34] [35] and the Fermi velocity $v_F = 2.2 \cdot 10^5$ m s$^{-1}$ [33] for Permalloy, we estimate the spin-flip time as $T_1 = v_F \text{l}_{sf}$ = 13 fs -37 fs. In the degenerate limit of $T_2 = 2T_1$ where spin-flip causes spin-decoherence, we estimate the maximum possible spin decoherence time as $T_2$ = 26 fs -74 fs, which bounds the fitted value we obtained for $\tau_{sc}$.

An alternative explanation is provided by the theory of lateral diffusion of spin-current generated by spin-pumping into an adjacent non-magnetic layer. However, the calculated increase in damping obtained by application of the theory in Refs [36] [37] to our micromagnetic simulation results is more than an order of magnitude smaller than what we observed.

Damage and/or oxidation at the sidewalls of a nanomagnet, which was potentially introduced during ion milling or after the patterning process, has been proposed as a source of enhanced damping [17]. To test this hypothesis, we performed micromagnetic simulations with enhanced damping at the nanomagnet edges modeled by $\alpha(y,z) = 0.0074 + \alpha' e^{-\left[\sqrt{(z/\varepsilon)^2 + y^2} - R\right]/\delta}$, where $\alpha'$ is the enhanced damping at the edge, $\delta$ is the decay length, $\varepsilon$ is the nanomagnet ellipticity, and $R$ is the length of the short axis. We used parameter values $\alpha' = 0.003$ and $\delta = 20$ nm. The decay length was chosen to match the zone of altered contrast in transmission electron microscope images of magnetic nanostructures [38], and $\alpha'$ was chosen such that the simulation results match the average measured damping values for the end and center modes of the 400 nm nanomagnets. We find that the nonuniform damping profile leads to negligible mode distortions relative to those obtained with uniform damping. The effective damping $\alpha_{eff}$ was determined by simulating swept-field FMR to determine $\Delta H$, and then using eq. (2) to extract $\alpha_{eff}$, with resultant values shown in Fig. 3(c). In the case of the 400 nm and 200 nm nanomagnets, the difference in the values of damping for the end- and center-modes is easily accommodated with such a spatial model of edge-enhanced damping: The end-mode is more localized near the edges, therefore $\alpha_{eff}$ is significantly enhanced for the end-modes. However, the model breaks down in the case of the 100 nm nanomagnets. While simulations predict that $\alpha_{eff}$ increases, the data clearly show that the damping for the 100 nm nanomagnet end-mode is significantly less than the end-mode damping for both the 200 nm and the 400 nm nanomagnets. Thus, edge damage fails to explain the observed trend for $\alpha$.

Therefore, we conclude that our measured values for $\alpha$ for discrete spin-wave eigenmodes in individual, isolated nanomagnets are well explained by the theory of non-local damping due to intralayer dissipative transverse spin-currents.

Acknowledgement
We would like to thank Y. Tserkovnyak, M. Schneider and M. Donahue for helpful discussions.

Fig. 1

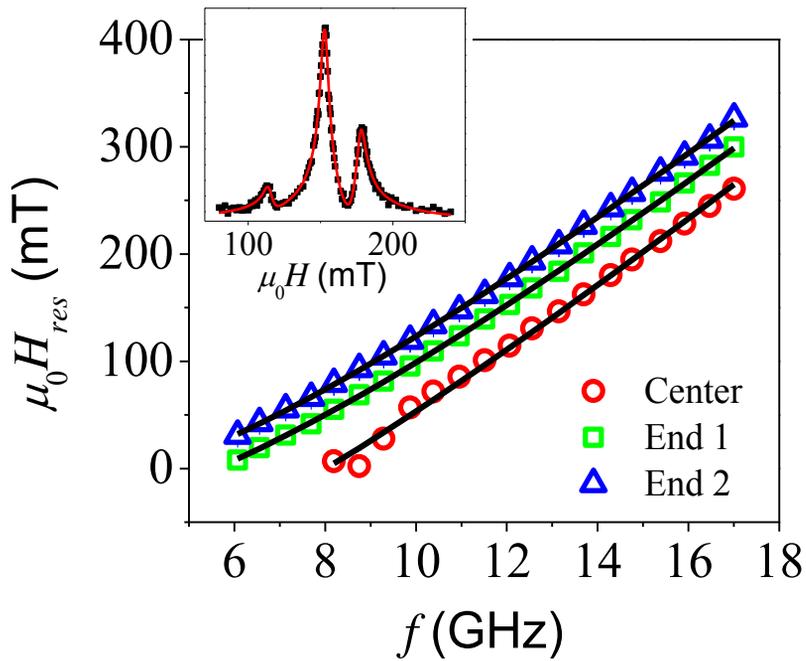

Fig. 1: The measured resonance fields of the three spin-wave modes for a 100 nm nanomagnet are shown. The center mode (red circles) has the lowest resonance field followed by the end-mode 1 (green squares) and end-mode 2 (blue triangles). The solid lines are fits to Eq. (1). The inset shows a spectrum obtained at 13.2 GHz. The red line is a fit to Eq. (1) in the SI.

Fig. 2

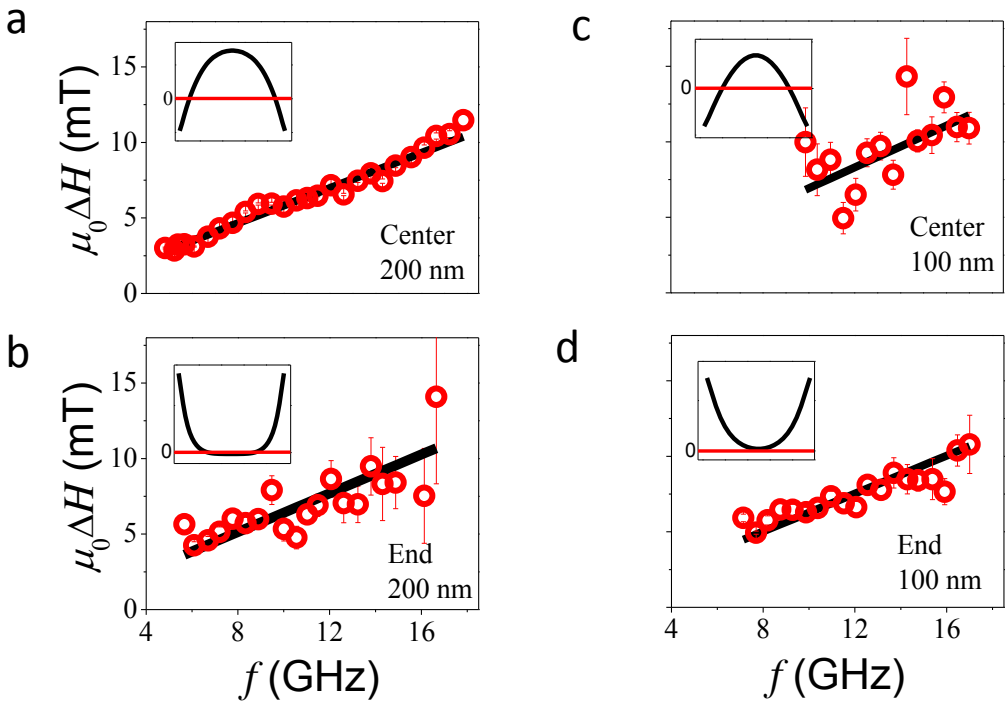

Fig. 2: Linewidths for the center-modes (a and c) and end-modes (b and d) for a 200 nm and a 100 nm nanomagnet. The insets show the mode profile along the long axis of the ellipsoid, as determined by micromagnetic simulations. The horizontal red line in the insets indicates zero amplitude.

Fig. 3

(a) Experimental Data

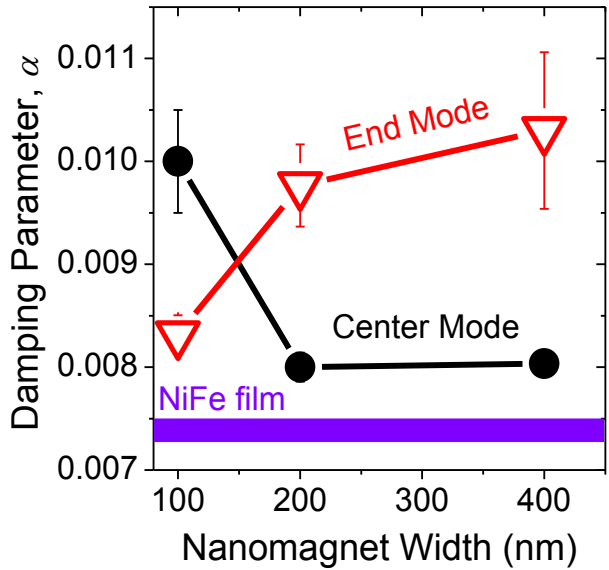

(b) Intralayer Spin Pumping Model

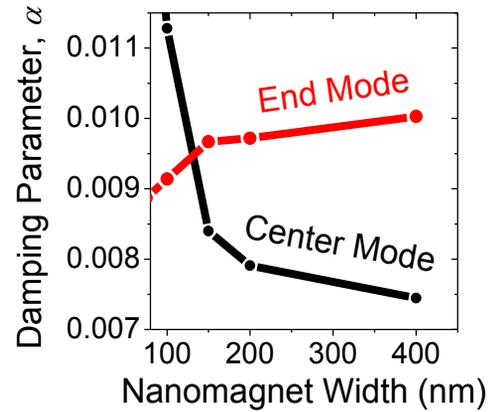

(c) Edge Damping Model

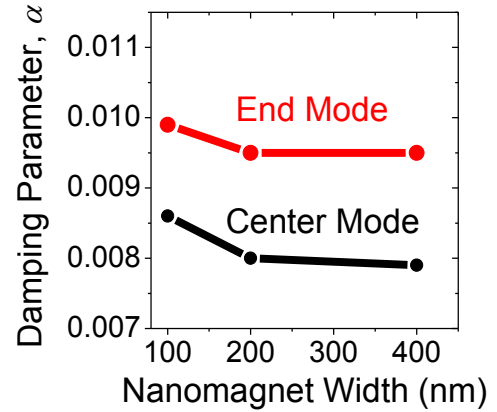

Fig. 3: (a) Experimental damping data: We plot the dependence of $\alpha$ on nanomagnet size for the different modes. The black circles (red triangles) are the average values of $\alpha$ for the end modes (center mode). The measured value for a 20 x 20 μm² square is marked in both with a purple bar, where the width of the bar indicates the measurement precision (b) Intralayer spin-pumping model: the solid red circles are the fitted values of $\alpha$ for the end-mode and the black circles for the center mode. $\tau_{sc}$ was the only fitting parameter. (c) Edge-enhanced damping model: red circles are the estimated values of $\alpha$ for the end mode and the black circles for the center mode.

## Heterodyne magneto-optical microwave microscope (H-MOMM):

In the H-MOMM, the sample is mounted with the nanomagnets positioned over the 100-μm-wide center conductor of a coplanar waveguide for the purpose of exciting FMR. Two tunable, single-frequency lasers (the "probe" and "LO") are used for the measurement. A microwave excitation is generated by mixing portions of the detuned probe- and LO-beams on a broadband photodiode, which is then amplified to 1 W and fed into the coplanar waveguide. Detection of the magnetization dynamics is achieved via the polar magneto-optic Kerr effect, whereby the polarization angle of linearly polarized light is rotated upon reflection in proportion to the perpendicular component of magnetization. The linearly polarized probe beam is focused onto a single nanomagnet, after which the back-reflected probe beam is passed through a polarization analyzer, then mixed with the LO beam. The mixing of the microwave-modulated probe beam with the LO beam generates a dc-signal on a photodiode with $V \sim |E_{LO}||E_{probe}|\phi_K$, where $E_{LO}$ and $E_{probe}$ are the electric fields of the LO beam and the reflected probe beam, respectively, and $\phi_K$ is the polarization rotation due to the magneto-optic Kerr effect activity of the nanomagnet. Thus the signal here is $V \sim |E_{LO}||E_{probe}|\phi_K$, whereas the signal is $V \sim |E_{probe}|^2$ in conventional Kerr microscopy. The linear dependence on $E_{probe}$ for heterodyne MOKE strongly enhances the SNR when measuring low-intensity magneto-optic signals in the case of specular reflection from nanomagnets with sizes significantly smaller than the focused laser-spot diameter.

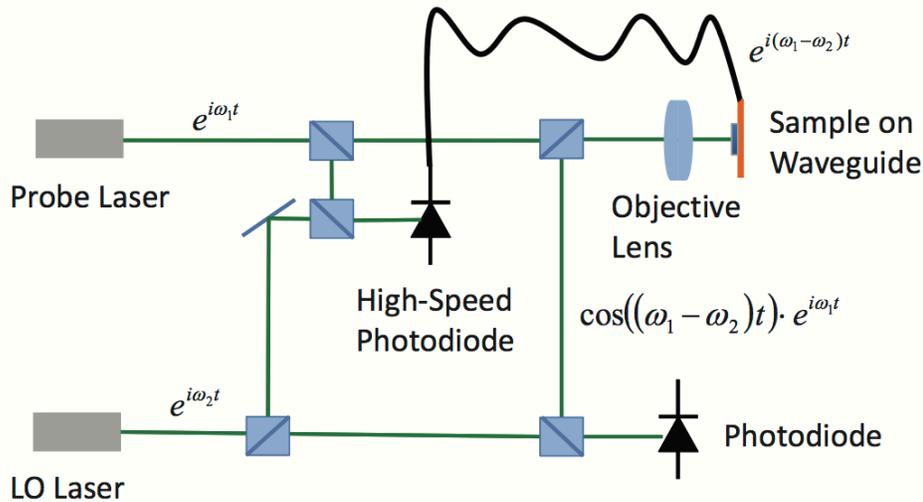

**Figure 1**: Simplified sketch of the heterodyne magneto-optical microwave microscope (H-MOMM). The probe laser and the local oscillator (LO) laser are detuned with respect to each other. The two laser beams are mixed on a high-speed photodiode and the resulting microwave signal is amplified and fed into a coplanar waveguide to excite ferromagnetic resonance. The magnetization of the nanomagnet is excited by the microwave field and precesses at the beat frequency of the two lasers. The linear polarized probe laser is focused onto the nanomagnet and its polarization is modulated by the precessing magnetization as a result of the polar magneto-optical Kerr effect (MOKE). The back reflected probe beam is mixed again with the LO beam and passes through a polarization analyzer. The mixed beams generate a DC signal on a photodiode, which is proportional to the MOKE rotation.

**Fitting procedure for the H-MOMM spectra:**

Fitting of all the field-swept spectra as a function of excitation frequency with full micromagnetic simulations is an impractical approach for the determination of $\alpha$ for each mode. Instead, we have chosen a simplified approach based upon the observation that the dependence of resonance field on frequency for all spin-wave modes is well fitted by the Kittel equation (eq. (1) in the manuscript). This permits us to characterize the data by effective stiffness fields used solely for the purpose of spectral fitting to extract the field-swept linewidth.

In a field-swept spectrum for the H-MOMM geometry with the *y*- and *z*-coordinates along the short and long axes, respectively, the detected component of the complex susceptibility tensor for a given spin-wave mode is approximated by that for a uniformly magnetized ellipsoid [1]:

$$\chi_{xy}(H) = \frac{2\pi M_s f}{|\gamma|\mu_0} \sum_i \frac{A^{(i)}}{\left(H+H_1^{(i)}\right)\left(H+H_2^{(i)}\right) - \left(\frac{2\pi f}{|\gamma|\mu_0}\right)^2 + i\Delta H\left(2H + H_1^{(i)} + H_2^{(i)}\right)}, \quad (1)$$

where $A^{(i)}$ is the complex amplitude of the *i*th mode, $\gamma = \frac{g\mu_B}{\hbar}$ is the gyromagnetic ratio, $g=2.073\pm0.009$ is the spectroscopic *g*-factor, $\mu_B$ is the Bohr magneton, $H$ is the external field, $\Delta H$ is the field-swept linewidth, $\mu_0 M_s=1.003\pm0.01$ T is the saturation magnetization, and $f$ is the frequency of the microwave field. $H_1^{(i)}$ and $H_2^{(i)}$ are the effective stiffness fields of the ith spin-wave mode, which include contributions from dipolar and exchange interactions. $M_s$ and $g$ for this $Ni_{80}Fe_{20}$ sample are obtained from H-MOMM measurements on the 20x20 μm² square. The measured spectra were fitted with the magnitude of eq. 1, see red line in fig. 2. It was necessary to treat the amplitude factors $A^{(i)}$ as complex fitting parameters to obtain a reasonable fit. The resonance field $H_{res}^{(i)}(f)$ for each mode was then fitted with the Kittel equation, see eq. 1 in the manuscript, over all measured frequencies to extract global values for $H_1^{(i)}$ and $H_2^{(i)}$. The fitted values of $H_1^{(i)}$ and $H_2^{(i)}$ were then used to refine the fits of the field-swept spectra to eq. (1) in order to improve the accuracy of the fitted value for the field-swept linewidth $\Delta H$. Micromagnetic simulations were used to confirm that this methodology is an accurate means of determining $\alpha$.

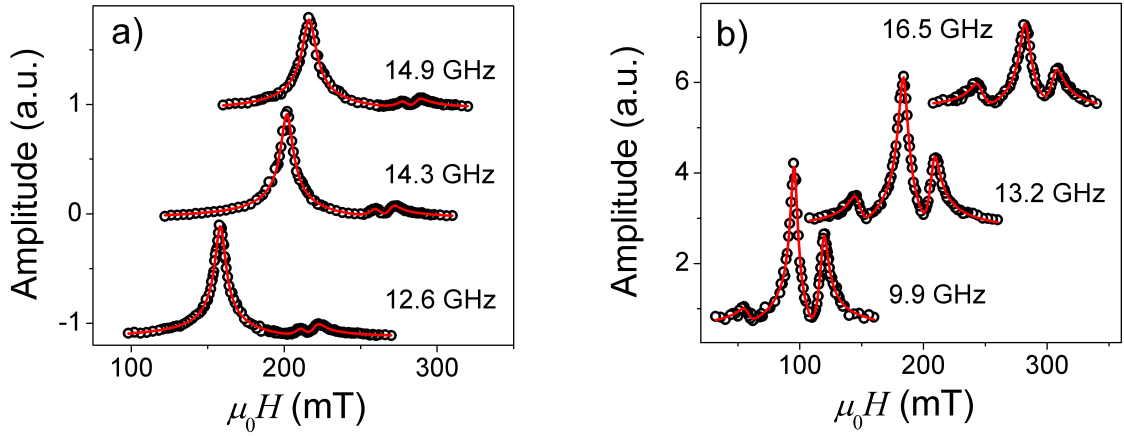

**Figure 2:** Spectra measured on (a) a 200 nm and (b) a 100 nm nanomagnets are shown. The red lines are fits of the data to Eq. 1. The mode with the lowest resonance field is the center-mode and the two modes with the higher resonance field are the non-degenerate end-modes.

**Intralayer damping:**

The intralayer spin-currents result in an additional nonlocal torque in the Landau-Lifshitz-Gilbert equation,

$$\partial_t \vec{m} = -\mu_0 |\gamma| \vec{m} \times \vec{H}_{eff} + \alpha \vec{m} \times \partial_t \vec{m} - \frac{|\gamma|}{M_s} \partial_i \vec{j}_i, \qquad (2)$$

where $\vec{m} = \vec{M}/M_s$ is the normalized magnetization, $\vec{H}_{eff}$ is the effective field, and $\vec{j}_i$ the spin-current-density flowing in the $i$th direction. The dissipative component of $\vec{j}_i$ that has components linear in excitation amplitude is given by [2]

$$\vec{j}_i = -\sigma_T \vec{m} \times \partial_i \partial_t \vec{m}, \qquad (3)$$

where $\sigma_T = (\hbar/2)^2 n_e \tau_{sc} / m^*$ is the transverse spin conductivity, $n_e$ is the conduction electron density, $m^*$ the effective mass, and $\tau_{sc}$ is the transverse spin scattering time, which can have contributions from momentum scattering, e-e interactions, as well as spin-orbit induced spin-flip/decoherence processes. Substituting (3) into (2) and dropping contributions $\propto \partial_i \vec{m} \times \partial_i \partial_t \vec{m}$, which are quadratic in the excitation amplitude, one obtains the additional damping due to transversal spin currents

$$\Delta\alpha = -\frac{\sigma_T|\gamma|}{M_s} \cdot \frac{\int_{Area} \delta m(\vec{x}) \nabla^2 \delta m(\vec{x}) d^2 x}{\int_{Area} |\delta m(\vec{x})|^2 d^2 x} = \frac{\sigma_T|\gamma|}{M_s} \cdot \frac{\int k^2 |\delta m(\vec{k})|^2 d^2 \vec{k}}{\int |\delta m(\vec{k})|^2 d^2 \vec{k}}, \quad (4)$$

where $\delta m(\vec{x})$ is the spatial profile of the spin-wave mode, and $\delta m(\vec{k})$ is the respective Fourier transform. The spatial profile $\delta m(\vec{x})$ for each spin-wave mode is obtained from micromagnetic simulations. We evaluate eq. (4) for each eigenmode. Because the 2-d Fast Fourier Transform over an ellipse embedded in a rectangular domain results in additional spectral components unrelated to the magnetization profile within the ellipse, we restrict the integration to a stripe down the middle of the ellipse that excludes the lateral edges. For the end mode, this results in a reasonable estimate for the damping, given that the amplitude of the end-mode is negligible outside of the considered region. In the case of the center-mode, the curvatures $\partial^2(\delta m(\vec{x}))/\partial x^2$ and $\partial^2(\delta m(\vec{x}))/\partial y^2$ are approximately constant or proportional to $\delta m(\vec{x})$, thus the additional damping is independent of the integration area and the integration over the stripe also results in a reasonable estimate for the damping.

In Fig. 3(a) of the manuscript we show the measured $\alpha$ and in (b) the best fit of the data to eq. (4), with the result $\tau_{sc}$ = 49 fs as the sole fitting parameter. We use $n_e = k_F^3/3\pi^2 = 3.9 \cdot 10^{28}$ m$^{-3}$ from the measured Fermi wavenumber $k_F = 1.05 \cdot 10^{10}$ m$^{-3}$ for the majority band in Permalloy [3] and the free electron mass for $m^*$.

**Damping parameter α for all the measured nanomagnets:**

| Short axis (nm) | Mode | α | Error |
|---|---|---|---|
| 400 | CM | 0.0080 | 0.0001 |
|  | EM1 | 0.0107 | 0.0004 |
| 400 | CM | 0.0081 | 0.0001 |
|  | EM1 | 0.010 | 0.001 |

| 200 | CM | 0.0084 | 0.0001 |
|---|---|---|---|
|  | EM1 | 0.0092 | 0.0003 |
|  | EM2 | 0.0094 | 0.0003 |
| 200 | CM | 0.0078 | 0.0001 |
|  | EM1 | 0.0112 | 0.0007 |
|  | EM2 | 0.0094 | 0.0003 |
| 200 | CM | 0.0078 | 0.0002 |
|  | EM1 | 0.0094 | 0.0003 |
|  | EM2 | 0.01 | 0.0004 |
| 100 | EM1 | 0.0079 | 0.0002 |
|  | EM2 | 0.0081 | 0.0002 |
| 100 | CM | 0.01 | 0.0005 |
|  | EM1 | 0.0077 | 0.0001 |
|  | EM2 | 0.009 | 0.0002 |
| 100 | EM1 | 0.0084 | 0.002 |
|  | EM2 | 0.0088 | 0.002 |

**Table 1**: Summary of the Landau-Lifshitz damping a for all the measured nanomagnets. The spin-wave modes are designated as center-mode (CM), end-mode 1 (EM1) and end-mode 2 (EM2). The error for $\alpha$ is one standard deviation, as determined from the fit to the data to Eq. 2 in the manuscript.